\documentclass[11pt]{article}

\usepackage{calc}
\usepackage[dvips]{graphicx}
\usepackage{xcolor}
\usepackage{epsfig}
\usepackage{amsmath,amsfonts,amssymb,amsthm}
\usepackage{verbatim}
\usepackage{psfrag}
\usepackage{bm}
\usepackage{bbm}
\usepackage[square,comma,sort&compress,numbers]{natbib}
\usepackage{color}
\usepackage{xcolor}
\usepackage{slashed}
\usepackage{upgreek}
\RequirePackage[colorlinks=true
  ,urlcolor=blue
  ,anchorcolor=blue
  ,citecolor=blue
  ,filecolor=blue
  ,linkcolor=blue
  ,menucolor=blue
  ,linktocpage=true
  ,pdfproducer=medialab
  ,pdfa=true
]{hyperref}

\usepackage[T1]{fontenc}

\usepackage{cancel}

\usepackage{epsf,epsfig}
\usepackage{graphics}
\usepackage[normalem]{ulem}

\def\d{\mathrm{d}}
\def\L{\mathcal{L}}
\def\i{\mathrm{i}}
\def\e{\mathrm{e}}

\begin{document}

\begin{flushright}
    {MITP-25-070\\
    TUM-HEP-1576/25}
\end{flushright}
\begin{center}
\large\bf\boldmath
Reply to ``Clearing up the Strong $CP$ problem''\\
\unboldmath
\end{center}


\begin{center}
{Wen-Yuan Ai$^a$, Bj\"orn~Garbrecht$^b$, Carlos Tamarit$^c$}\\
\vskip0.3cm
{\it\footnotesize $^a$Marietta Blau Institute for Particle Physics, Austrian Academy of Sciences,\\ Dominikanerbastei 16, A-1010 Vienna, Austria\\[1mm]
$^b$Physik-Department T70, James-Franck-Stra{\ss}e,\\
Technische Universit\"at M\"unchen, 85748 Garching, Germany\\[1mm]
$^c$PRISMA+ Cluster of Excellence \& Mainz Institute for Theoretical Physics,\\
Johannes Gutenberg-Universität Mainz, 55099 Mainz, Germany
}
\end{center}

\begin{abstract}
The conservation of $CP$ in QCD  has been shown to follow from a careful treatment of the path integral and canonical quantization in Refs.~\cite{Ai:2020ptm,Ai:2024vfa}. Here, we refute the critique of these results put forth in  Ref.~\cite{Benabou:2025viy}.   First, using the quantum rotor as an analogue of QCD, it is argued in Ref.~\cite{Benabou:2025viy} that the topological susceptibility vanishes when using the limiting procedure of Ref.~\cite{Ai:2020ptm}. When translated to QCD, this would contradict the observed $\eta^\prime$-mass. We show that this is not the case because the susceptibility is defined from the vacuum correlator of the topological charge density, which for the rotor is just fixed by the canonical commutation relation. The latter does not depend on the disputed order of limits. Second, it is suggested in Ref.~\cite{Benabou:2025viy} that $CP$ violation in QCD can be established by considering the low-energy effective theory alone. We show that here the argument relies on assuming from the start choices of couplings that lead to $CP$ violation but are not of the most general form allowed by spurion analysis. No valid reason is given for why allowed choices leading to $CP$ conservation, that match the computation of ultraviolet correlators as shown in Refs.~\cite{Ai:2020ptm,Ai:2024vfa}, would be inconsistent.
\end{abstract}

\paragraph{Summary and context.}
In Refs.~\cite{Ai:2020ptm,Ai:2024vfa} it is concluded that the strong interactions are $CP$ conserving by using path integral methods as well as canonical quantization. The key point in~Ref.~\cite{Ai:2020ptm} is that when calculating correlation functions in Yang--Mills theory from the Euclidean path integral at zero temperature, there is no physical reason for imposing boundary conditions that vanish up to gauge transformations on \emph{finite} subvolumes of spacetime. As a consequence, the spacetime volume must be taken to infinity before integer topological sectors can be assumed. The correlation functions then turn out to be $CP$ even, irrespective of the rephasing invariant angle $\bar \theta$~\cite{Ai:2020ptm}. From the point of view of canonical quantization, the deciding issue in Ref.~\cite{Ai:2024vfa} is that, when treating \emph{all} gauge transformations as a redundancy, consistency with the group structure of quantum chromodynamics (QCD) leaves no room for physical $CP$-odd phases.

The above conclusions are criticized in Ref.~\cite{Benabou:2025viy}, which contains no direct objection to the aforementioned order of limits or justification of the opposite one. Neither does it present counterclaims concerning the treatment of gauge transformations in canonical quantization. However, it contains arguments based on the topological susceptibility $\chi$ and chiral effective Lagrangians aiming to demonstrate that taking the spacetime volume to infinity first would lead to inconsistencies. Here, we explain why these arguments are unsound.

First, it is claimed that taking the spacetime volume to infinity before fixing the topological sectors does not accord with the observation of a heavy $\eta^\prime$-meson. Since the $\eta^\prime$-mass is a signal of the anomalous nonconservation of $U(1)_{\rm A}$, this appears surprising because of the fact that the divergence of the chiral current is a local identity that does not depend on boundary conditions implied by the different orders of limits. Furthermore, as discussed in Ref.~\cite{Ai:2020ptm}, the QCD predictions obtained by taking the infinite volume first can be matched to a chiral Lagrangian which, while $CP$ conserving, predicts a massive $\eta^\prime$-meson. Yet, Ref.~\cite{Benabou:2025viy} refers to a calculation from Ref.~\cite{Albandea:2024fui}, based on a quantum-mechanical rotor, that seems to suggest that $\chi$ vanishes when taking the volume to infinity first. If correct, this would imply that the $\eta^\prime$-mass could not be explained through the Witten--Veneziano relation~\cite{Witten:1979vv,Veneziano:1979ec} in fixed topological sectors and infinite spacetime volume.

However, for the rotor model from Ref.~\cite{Albandea:2024fui}, $\chi$ does in fact not vanish. For a rotor with angle $\phi(\tau)$ and $\tau$ being the Euclidean time variable,  the topological charge density $q(\tau)$ is defined as $\frac{1}{2\pi}\frac{\d\phi}{\d\tau}$. The topological susceptibility is then $\langle q(\tau)q(\tau^\prime)\rangle=\frac{1}{4\pi^2}\frac{\d}{\d\tau}\frac{\d}{\d\tau'}\langle 0| T \phi(\tau)\phi(\tau^\prime)|0\rangle=\delta(\tau-\tau^\prime)/(4\pi^2 m)$, where $m$ is the moment of inertia of the rotor. This correlation function is fully local and can be seen to simply follow from the canonical commutation relations. Hence, it is not even of a proper topological origin. Computing the topological susceptibility $\chi=\int{\rm d}\tau\,\langle q(\tau)q(\tau^\prime)\rangle=1/(4\pi^2 m)$, it is evident that it is independent of whether the integration interval is finite or infinite, as long as we pick up the coincident point of the correlator. Thus, there is no conflict with the topological susceptibility and the $\eta^\prime$-mass, contrary to the statements found in Ref.~\cite{Benabou:2025viy}. Further below, we roll out some more details on the correlation functions on infinite time intervals and pin down why the results presented in Refs.~\cite{Albandea:2024fui,Benabou:2025viy} for infinite time intervals cannot be identified with $\chi$.

Second, regarding the effective chiral Lagrangian for the meson field matrix $U$, its couplings are constrained by the anomalous chiral symmetries of QCD. These can be rephrased in terms of a spurious symmetry that should leave the effective theory invariant. Under this symmetry, the meson field matrix, the phases $\alpha_i$ of the masses of the $N_f$ quarks and the $\theta$-parameter (i.e. the angle in front of the topological term in Yang--Mills theory) transform as follows: $U\rightarrow \e^{2\i\beta} U,\,\alpha_i\rightarrow \alpha_i-2\beta,\,
\theta\rightarrow\theta+2N_f\beta$. The above transformation properties allow that the chiral condensate $\langle U\rangle$ can take a phase $\det\langle U\rangle\propto \e^{\i\xi}$, where either $\xi=\theta$ or $\xi=-\sum_i\alpha_i\equiv-\bar\alpha$. The option  $\xi=-\bar\alpha$ leads to no $CP$-violating effects and follows from matching with the correlations obtained in the fundamental theory by taking the spacetime volume to infinity first~\cite{Ai:2020ptm}. The other option, $\xi=\theta$, is selected by the correlators computed in the alternative order of limits. Since both options rely on the matching to well-defined correlation functions arising from the respective order of limits, it is not possible to determine the physical value of $\xi$ just from the effective field theory Lagrangian without matching. Yet, in Ref.~\cite{Benabou:2025viy} it is argued for the opposite, but based on effective Lagrangians that do not display the most general form under the spurious symmetries to start with, so that the reasoning is circular and inconclusive. The particular details of the arguments in Ref.~\cite{Benabou:2025viy} concerning effective chiral Lagrangians shall be addressed below.

\paragraph{Topological susceptibility.}
In the quantum rotor, one can talk about topological sectors labelled by integers $\Delta n$ when imposing periodic boundary conditions for the angle: $\phi(\tau_f)-\phi(\tau_i)=2\pi\Delta n$. The topological charge $\Delta n$ is then simply given as $\Delta n = \int \d\tau\, {\dot \phi(\tau)}/{2\pi}=\int \d\tau\, q(\tau).$
It is argued in Refs.~\cite{Benabou:2025viy,Albandea:2024fui} that the topological susceptibility cannot be obtained by first computing correlation functions within individual topological sectors in the infinite-volume limit and then summing over the sectors. This is the order of limits appropriate to obtain vacuum correlations~\cite{Ai:2020ptm},  which can also be understood to follow from defining the saddle point expansion of the path integral by means of a continuous deformation of the original integration contour~\cite{Ai:2024cnp}.\footnote{It is stated in Ref.~\cite{Benabou:2025viy} that the results in Refs.~\cite{Nakamura:2021meh,Schierholz:2024var,Schierholz:2025tns} also rely on this order of limits. This is not the case. Rather, it is claimed in these papers that the strong coupling is screened at large distances unless the $CP$-odd rephasing-invariants in the Lagrangian are zero.} The assertion of Ref.~\cite{Benabou:2025viy} is based on a calculation of the susceptibility for a quantum-mechanical rotor that yields zero for each fixed topological sector on an infinite time interval. Then, it trivially gives zero when summing over topological sectors. In the following, we explain how such a result arises and why it does not hold.

In quantum mechanical analogies to QCD,  $q(\tau)$ should be compared with the topological charge density $F\widetilde F$, with $F$ being the non-Abelian field strength and $\widetilde F$ its Hodge dual. The topological susceptibility $\chi$ is derived from a correlator with two insertions of $q(\tau)$ in the Euclidean path integral.
Following the standard definitions in the literature~\cite{Witten:1979vv,Shifman:1979if}, the starting point can be
\begin{align}\label{eq:chi_path_integral}
  \chi=  \lim_{k\rightarrow0}\int \d\tau\, \e^{\i k \tau}\left.\langle  q(\tau) q(0)\rangle\right|_{\theta=0}=\int \d\tau\int{\cal D}q(\tau)\, q(\tau) q(0) \e^{-\left.S_{\rm E}[q(\tau)]\right|_{\theta=0}}\,,
\end{align}
where $S_{\rm E}$ is the Euclidean action with a total derivative term $-{\rm i}\int\d\tau\, \theta q$.
Note that such a definition involving a single temporal integration is also quoted in Ref.~\cite{Benabou:2025viy}. Given the infinite time interval, this path integral definition is equivalent to
\begin{align}\label{eq:chi_canonical_quantization}
     \chi=  \frac{1}{4\pi^2}\int \d\tau\,\frac{\d}{\d\tau}\frac{\d}{\d\tau'}\,\left.\langle 0| T \phi(\tau) \phi(0)|0\rangle\right|_{\theta=0}\,,
\end{align}
where $|0\rangle$ represents the ground state, while $T$ is the usual time-ordering operator, with $T  \phi(\tau) \phi(\tau^\prime)=\Theta(\tau-\tau^\prime)\phi(\tau) \phi(\tau^\prime)+\Theta(\tau^\prime-\tau)\phi(\tau') \phi(\tau)$ and $\Theta$ being the Heaviside function. Equation~\eqref{eq:chi_path_integral} immediately follows from Eq.~\eqref{eq:chi_canonical_quantization} when the correlator $\langle 0| T \phi(\tau) \phi(0)|0\rangle$ is represented as a path integral with an insertion of $\phi(\tau) \phi(0)$ and one subsequently uses $q(\tau)=\frac{1}{2\pi}\frac{\d\phi}{\d\tau}$. Note that the time derivatives in Eq.~\eqref{eq:chi_canonical_quantization}  cannot be taken inside the correlator, past the $T$ operation, as the latter is time-dependent.\footnote{In the present article, we show the position of temporal derivatives when these explicitly act on operators, i.e. in formulae involving the states $|0\rangle$. In all other expressions, the derivatives are understood to act from outside on the expectation values. This convention is different from the one in Refs.~\cite{Witten:1979vv,Shifman:1979if}, where the derivatives are always acting from outside, also in expectation values of operators such as  $\langle 0| T \, F\widetilde F(\tau)\, F\widetilde F(0)|0\rangle$. That interpretation is consistent with the derivation of these expectation values from the path integral, which implies derivatives acting from outside.}
In view of the definition~\eqref{eq:chi_canonical_quantization} in terms of the ground state correlation function, we can compute the integrand in canonical quantization. Analytically continuing to Minkowski time $t$, it is straightforward to obtain
\begin{align}
\label{corr:operators}
\frac{1}{4\pi^2}\frac{{\rm d}}{{\rm d}t} \frac{{\rm d}}{{\rm d}t^\prime}\,\left\langle 0\left| T \,\phi(t)\,\phi(t^\prime) \right|0\right\rangle|_{\theta=0}=\frac{1}{4\pi^2}\left.\left\langle 0\left| T\, \frac{{\rm d}}{{\rm d} t}\phi(t)\,\frac{{\rm d}}{{\rm d}t^\prime}\phi(t^\prime)|\right|0 \right\rangle\right|_{\theta=0}+\frac{{\rm i}}{4\pi^2m} \delta(t-t^\prime)\,. 
\end{align}
This result follows from the definition of the time-ordering operation given above, the fact that ${\d\phi/\d t}$ commutes with the Hamiltonian and is thus time-independent and the equal-time canonical commutation relation $[\phi(t), m \,\d{\phi}(t)/\d t]={\rm i}$.
The first term on the right-hand side of Eq.~\eqref{corr:operators} vanishes for the ground state with zero angular momentum $\d{\phi}/\d t$. For this state, the correlation function is therefore proportional to the Dirac-$\delta$, and after analytical continuation of the integration in Eq.~\eqref{eq:chi_canonical_quantization} to ordinary time, we obtain the topological susceptibility
\begin{align}
\label{topo:susc}
\chi=-{\rm i}\frac{1}{4\pi^2}\int\limits_{-\infty}^\infty{\rm d}t\, \frac{{\rm d}}{{\rm d}t} \frac{{\rm d}}{{\rm d}t^\prime}\,\left.\langle T \phi(t)\,\phi(t^\prime) \rangle\right.=\frac{1}{4\pi^2 m}\,.
\end{align}
The question is whether this result can be reproduced from a path integral computation within a fixed topological sector on an infinite Euclidean time interval. Since the Dirac-$\delta$ function arises as a consequence of the canonical commutation relation, a negative answer would raise a more fundamental concern: Even a free point particle on a line might then fail to be faithfully quantized by the Euclidean path integral in the infinite-time limit, regardless of the choice of unphysical boundary conditions $\phi(\pm\infty)$.

In a Euclidean path integral setting with a time integration interval given by $\beta$, the ground state correlator $\langle 0| T \,\phi(\tau)\,\phi(\tau^\prime)|0\rangle$ corresponds to the $\beta\rightarrow\infty$ limit of the Green's function $G(\tau,\tau^\prime;\beta)$, subject to the appropriate boundary conditions. We take the latter to be $\phi(\pm\beta/2)=\phi_\pm$. While a sector of integer topological charge $\Delta n$ corresponds to choosing periodic boundary conditions $\phi_+-\phi_-=2\pi\Delta n$, we note that it is not necessary to impose specific boundary conditions to recover the ground state correlator. The Green's function for the operator $-m\frac{{\rm d}^2}{{\rm d}\tau^2}$ is found straightforwardly,
and we obtain~\cite{2025quantum-rotor}
\begin{align}
\label{correlator:F}
 F(\tau,\tau^\prime;\beta):=\frac{1}{4\pi^2}\frac{\rm d}{{\rm d}\tau}\frac{\rm d}{{\rm d}\tau^\prime}G(\tau,\tau^\prime;\beta)=\frac{1}{4\pi^2 m}\delta(\tau-\tau^\prime)-\frac{1}{4\pi^2 \beta m}+\frac{1}{4\pi^2}\left(\frac{\phi_+-\phi_-}{\beta}\right)^2\,.
\end{align}
We observe that, $F(\tau,\tau^\prime) \equiv\lim_{\beta\to\infty} F(\tau,\tau^\prime;\beta)$---the object that is found for a fixed topological sector with an infinite time interval---when analytically continued to real time, is consistent with the correlation function for the zero-angular momentum state in Eq.~(\ref{corr:operators}). At the level of correlation functions, the limit $\beta\to\infty$ is thus well-defined, and it must be taken before any subsequent temporal integration over an infinite interval in order to project on the ground state correlator~(\ref{corr:operators}) and to avoid artifacts due to contributions from excited states. Consequently, the susceptibility follows for a fixed sector on an infinite Euclidean time interval as
\begin{align}
\chi=\int\limits_{-\infty}^\infty{\rm d}{\tau}\, F(\tau,\tau^\prime)=\frac{1}{4\pi^2 m}\,.
\end{align}
This nonzero value for a fixed sector, which holds for all sectors, contradicts the vanishing result in Ref.~\cite{Albandea:2024fui} quoted in Ref.~\cite{Benabou:2025viy}. In these references, the topological susceptibility in an individual sector is evaluated as $\chi_{\Delta n,\beta}=\Delta n^2/\beta$, which goes to zero when sending $\beta\rightarrow\infty$ for fixed $\Delta n$. 

To clarify the origin of this discrepancy, we can understand how one may arrive at the zero result from the Green's function given in Eq.~\eqref{correlator:F}. If, instead of evaluating $\lim_{\beta\to\infty} F(\tau,\tau^\prime;\beta)$ first before integrating over an infinite time, one evaluated the integrand for finite $\beta$ and took $\beta\rightarrow\infty$ in the end, one would get
\begin{align}
    \int_{-\beta/2}^{\beta/2}\d\tau\, F(\tau,0;\beta)= \frac{(\phi_+-\phi_-)^2}{4\pi^2\beta}\,.
\end{align}
For periodic boundary conditions $\phi_+-\phi_-=2\pi\Delta n$ one would  recover the result $\Delta n^2/\beta$ used in Refs.~\cite{Benabou:2025viy,Albandea:2024fui}. However, with this procedure, the result does not correspond to the integral of a vacuum correlator as in the fundamental definition of Eq.~\eqref{eq:chi_canonical_quantization} because $F(\tau,0;\beta)$ only matches the vacuum correlator for $\beta\rightarrow\infty$. Therefore, the zero susceptibility quoted in Refs.~\cite{Benabou:2025viy,Albandea:2024fui} is a consequence of not using its correct definition in the ground state with an infinite time interval.  Once the definition in terms of the ground state correlator in Eq.~\eqref{eq:chi_canonical_quantization} is used, the result is nonzero. When extrapolating this result to QCD, the contradictions purported in Ref.~\cite{Benabou:2025viy} with the current algebra results from Refs.~\cite{Witten:1979vv,Veneziano:1979ec} and the mass of the $\eta^\prime$-meson disappear. We shall note that the former results are derived in terms of ground state correlators, and that is why one should use a definition of the topological susceptibility in terms of the same.

To avoid confusion, we note that in Ref.~\cite{Ai:2020ptm} by the present authors, the topological susceptibility is also derived based on $\Delta n^2/\beta$, which amounts, as seen here, to a definition different from the one corresponding to Eq.~(\ref{eq:chi_canonical_quantization}). It is then argued that the nonvanishing $\eta'$-mass, which follows from the explicit breaking of the chiral symmetry by instanton-induced interactions (`t~Hooft vertices), can still be linked to a nonzero topological susceptibility defined in subvolumes. However, as explained here, it turns out that this is not necessary when adhering to the primary definition corresponding to Eq.~(\ref{eq:chi_canonical_quantization}). Just as it is the case for the rotor, the corresponding definition should still give a nonvanishing susceptibility for QCD under the limiting procedure used in Ref.~\cite{Ai:2020ptm}. Note also Ref.~\cite{Kaplan:2024ezz} for recent progress in understanding the $\eta^\prime$-mass in a fixed topological sector.

\paragraph{Effective chiral Lagrangian.}

It is further argued in Ref.~\cite{Benabou:2025viy} that the results in Refs.~\cite{Ai:2020ptm,Ai:2024vfa,Ai:2024cnp} are inconsistent with chiral perturbation theory. This is done by asserting an effective Lagrangian that implies $CP$-violating effects but amounts to a restriction of the most general form allowed by spurion analysis. If there were valid reasons for the claimed restriction, $CP$ violation could be deduced without a direct derivation from the underlying topological effects in the fundamental QCD description, with which we disagree. Reference~\cite{Benabou:2025viy} presents three variants of the argument, each relying on implicit or ad hoc assumptions that enforce the restricted form of the interactions, which do not appear to have a justification on their own.

First, Ref.~\cite{Benabou:2025viy} argues that one can derive $CP$-violating effects based {\it solely} on the two leading terms in the effective chiral Lagrangian,
\begin{align}
    \L^{\rm EFT}\supset \frac{f_\pi^2}{4} {\rm Tr}(\partial_\mu U) \partial^\mu U^\dagger + \frac{f_\pi^2 B_0}{2}{\rm Tr}(MU+U^\dagger M^\dagger)\,.
\end{align}
For simplicity, let us consider the two-flavour case. Following Ref.~\cite{Benabou:2025viy}, we work in the basis where the $\theta$ angle has been rotated to the quark mass matrix and write $M={\rm diag}(m_u \e^{\i\bar\theta/2},m_d \e^{\i\bar\theta/2})$\footnote{For the more general case, see Ref.~\cite{Ai:2024cnp}.} where $\bar{\theta}=\theta+\bar{\alpha}$. To obtain $CP$ violation, the added assumption is that the expectation value of the determinant of the unitary matrix $U$ is equal to \emph{one}. Given the identification of $U$ with fermion bilinears $\bar\psi P_{\rm R} \psi$,\footnote{Our convention for the quark mass matrix $M$ is chosen as $\L_{\rm QCD}\supset -\bar\psi M P_{\rm R} \psi$. When comparing with the notation used in Ref.~\cite{Srednicki:2007qs}, our $M$ and $U$ differ by a Hermitian conjugation.} this amounts to imposing ad hoc restrictions on the phases of the quark condensate. However, $\det U=1$ is not necessarily the case, and the phase of the quark condensate must be determined by some extra argument.

Such an argument may involve adding a term that reflects the anomalous breaking of the chiral symmetry at the effective theory level. One possible form of it is
\begin{align}\label{eq:det_terms}
    \L_{\rm det}^{\rm EFT}= |\lambda| \e^{-\i\xi} f_\pi^4\det U +|\lambda| \e^{\i\xi} f_\pi^4\det U^\dagger\,.
\end{align} 
By spurion analysis the two options for $\xi$ discussed in Refs.~\cite{Ai:2020ptm,Ai:2024cnp} are
\begin{align}
\label{eq:options}
    \xi=\begin{cases}
        0 \\ -\bar\theta
    \end{cases} \quad (\text{in the basis where the topological angle $\theta$ is rotated to $M$})\,.
\end{align}
The first one leads to $CP$-violating effects while the second one does not. Setting aside the question of which option is correct, we now show that the phases in the quark condensate are correlated with $\xi$. Hence, imposing an ad hoc constraint on the determinant of $\langle U \rangle$ is not valid, as it is equivalent to an ad-hoc constraint on $\xi$. Any derivation of strong $CP$ violation in chiral perturbation theory without considering the determinant term implicitly relies on the assumption $\xi=0$ in the present basis.

Parametrizing
\begin{align}
    \langle U\rangle= \begin{pmatrix}
        \e^{\i \varphi_u} & 0\\
        0 & \e^{\i \varphi_d}
    \end{pmatrix}\,,
\end{align}
leads to a potential of the form
\begin{align}
    V(\varphi_u,\varphi_d)= -f_\pi^2 B_0 \left[m_u \cos\left(\frac{\bar\theta}{2}+\varphi_u\right)+m_d\cos\left(\frac{\bar\theta}{2}+\varphi_d\right)\right]-2|\lambda|f_\pi^4 \cos(\xi-\varphi_u-\varphi_d)\,.
\end{align}
One can then use the conditions $\partial V/\partial\varphi_u=\partial V/\partial \varphi_d =0$ to obtain $\varphi_u$, $\varphi_d$. In the limit $|\lambda|\gg B_0 m_d/f_\pi^2$ and assuming that $\xi+\bar\theta$ is in the first quadrant, we obtain
\begin{subequations}
\begin{align}
    &\sin(\xi-\varphi_u-\varphi_d)\,\approx\,0\quad \Rightarrow\quad  \xi\,\approx\,\varphi_u+\varphi_d\,, \label{eq:xi-varphiu-varphid}\\
    &m_u\,\sin\left(\frac{\bar{\theta}}{2}+\varphi_u\right)=m_d\,\sin\left(\frac{\bar{\theta}}{2}+\varphi_d\right)\,\approx\,\frac{\sin(\xi+\bar\theta)}{\sqrt{\frac{1}{m_u^2}+\frac{1}{m_d^2}+\frac{2\cos(\xi+\bar\theta)}{m_um_d}}}\,.
\end{align}    
\end{subequations}
Now we see that only for $\xi=0$ one can write $\varphi_u=-\varphi_d=\phi$, which is the form of $\langle U\rangle$ with determinant \emph{one} used in Ref.~\cite{Benabou:2025viy}. For other values of $\xi$, one could write $\varphi_u=\xi/2 +\phi$, $\varphi_d=\xi/2-\phi$. Precisely due to the additional phase $\xi$ in $\langle U\rangle$, one cannot conclude $CP$ violation from the following terms, that are consistent with the spurious symmetries and involve the isospin doublet of nucleons $N$: 
\begin{align}
     -c_1 \bar N (M^\dagger P_{\rm L} + M P_{\rm R} )N - c_2 \bar N (U M U P_{\rm L} + U^\dagger M^\dagger U^\dagger P_{\rm R} )N\,.
\end{align}
Following through the analysis of Ref.~\cite{Srednicki:2007qs}, after a field redefinition $N=(u^\dagger P_L+u P_R){\cal N}$ with $U=u^2$ and ${\cal N}$ the doublet of physical neutron and proton eigenstates $n$ and $p$, one can see that this term leads to $\L_{\pi p n}\sim (\bar\theta+\xi) \pi^+ \bar{p} n$, which vanishes for the second option in Eq.~\eqref{eq:options} and where $\pi^+$ is the charged pion. Note that while it is claimed in Ref.~\cite{Benabou:2025viy} that the effective Lagrangian term
$
    -c_3 {\rm Tr} [MU+M^\dagger U^\dagger]\bar{N} (U P_L+ U^\dagger P_R)N
$
contributes to the neutron dipole moment when $\xi=0$, this is not the case because the same field redefinition as above removes the $CP$-odd interactions involving  charged pions and nucleons. As field redefinitions leave the physics invariant, the electric dipole moment of the neutron cannot receive any contribution from $c_3$ at one-loop order. Note that one can also derive the $\eta^\prime$-mass (in the two-flavour case) as\footnote{There is a typo in Ref.~\cite{Ai:2024cnp}.} $m_{\eta^\prime}\approx 8|\lambda|f_\pi^2 \cos(\xi-\varphi_u-\varphi_d)$, whose value does not depend on the options for $\xi$ due to Eq.~\eqref{eq:xi-varphiu-varphid}. 

Second, Ref.~\cite{Benabou:2025viy} notes that if the second option for $\xi$ is correct, the effective Lagrangian has a spurious symmetry under $\psi_i\rightarrow \, \e^{\i\beta\gamma^5} \psi_i$ and (spurious) $M\rightarrow  \e^{-2\i\beta} M$. It is then stated that this is not consistent with QCD at the fundamental level because it is not a symmetry there. However, no reason for this alleged inconsistency is provided. Presumably, the argument is that under the field redefinition $\psi_i \to \e^{\i\beta\gamma^5} \psi_i$, the QCD \emph{partition function} is not invariant because the chiral anomaly induces a shift in $\theta$, and therefore the effective chiral \emph{Lagrangian} should also fail to be invariant. This conclusion does not follow. In the present case, the $\theta$-dependence of the partition function enters through a normalization factor that affects all correlators equally: $Z(\theta)=N(\theta) Z(\theta=0)$, as shown in Ref.~\cite{Ai:2020ptm}. As the normalization factor drops out from observables, physics is not sensitive to changes in $\theta$. In fact, constructing an effective action from the physical one-particle-irreducible correlators, which are independent of the normalization factor, one gets a result invariant under the transformations  $\psi_i\rightarrow  \e^{\i\beta\gamma^5}\psi_i$ and $M\rightarrow \e^{-2\i\beta} M$. At low energies, this effective action should be matched to the chiral Lagrangian, which should exhibit the same symmetry, as is indeed the case for the chiral Lagrangian with the second option for $\xi$ from Eq.~(\ref{eq:options}). Hence, there is no contradiction between such an effective Lagrangian and the properties of the QCD partition function.

Third, Ref.~\cite{Benabou:2025viy} uses the effective Lagrangian in the large-$N_c$ limit~\cite{DiVecchia:1980yfw} (in the general basis for the chiral phases)
\begin{align}
\label{eq:Vecchia-Veneziano}
    \L^{\rm EFT} \supset &\frac{f_\pi^2}{4} {\rm Tr} (\partial_\mu U)\partial^\mu U^\dagger +\frac{f_\pi^2 B_0}{2} {\rm Tr} (MU+U^\dagger M^\dagger)+\frac{\i}{2} q(x){\rm Tr}\left[\log U-\log U^\dagger\right]+\frac{N_c}{a f_\pi^2} q^2(x)-\theta q(x)\,,
\end{align}
where $q(x)$ is an auxiliary field. Once again, this is not the only allowed way the phases may enter the effective theory. By requiring the effective Lagrangian to be invariant under the spurious transformations
\begin{align}
    \psi_i\rightarrow \e^{\i\beta\gamma^5}\psi_i\,, \quad \theta\rightarrow \theta+ 2 N_f\beta\,,\quad M\rightarrow \e^{- 2\i \beta} M\,,
\end{align}
the last term in Eq.~\eqref{eq:Vecchia-Veneziano} allows for a more general form as
\begin{align}
    -\theta q(x)\rightarrow -\xi q(x)\,,
\end{align}
with the two options 
\begin{align}
\label{eq:options2}
    \xi=\begin{cases}
        \theta \\ -\bar{\alpha}
    \end{cases} \quad (\text{in the general basis where } M={\rm diag}(m_1 e^{\i\alpha_1},...,m_{N_f} \e^{\i\alpha_{N_f}}))\,.
\end{align}
When working in the basis where $M$ is real, then one gets
\begin{align}
\label{eq:options3}
    \xi=\begin{cases}
        \bar{\theta} \\ 0
    \end{cases} \quad (\text{in the basis where $M$ is real)}\,.
\end{align}
It is the first option that is assumed in Refs.~\cite{DiVecchia:1980yfw,Benabou:2025viy}. To determine which option is the correct one, one must match the effective theory with the phase in the correlation functions computed from fundamental QCD. This is what is done in Ref.~\cite{Ai:2020ptm}, which in the general basis for the chiral phases results in $\xi=-\bar\alpha$, the second option in Eq.~\eqref{eq:options2}. Note that for either option, one gets $m^2_{\eta'}$ scaling as $1/N_c$. Furthermore, using $\xi=-\bar\alpha$ in the general basis and integrating out $q(x)$ leads to a Lagrangian for $U$ whose $\log U$-dependent terms match the expansion of the $\det U$ terms in Eq.~\eqref{eq:det_terms} with $\xi=-\bar\alpha$, making a connection with the chiral Lagrangian derived in Ref.~\cite{Ai:2020ptm}.

Note that one can identify $q(x)=\frac{g^2}{64\pi^2}G(x)^{\mu\nu}\widetilde{G}(x)_{\mu\nu}$ from anomalous current conservation. While such a point is not explicitly brought forward in Ref.~\cite{Benabou:2025viy}, this may seem to suggest that an effective term $\theta q(x)$ is just what matches the situation in the fundamental theory. However, it is a general feature of effective field theories that there is no equivalence between parameters in the fundamental theory and their corresponding version in the effective theory. When integrating out fields across thresholds, the parameters in the effective theory absorb quantum corrections, and their value is fixed not by matching with the ultraviolet (UV) Lagrangian, but by matching correlators computed with both the fundamental and effective theories. This generically leads to departures between fundamental and effective parameters. The Lagrangian of Eq.~\eqref{eq:Vecchia-Veneziano} assumes an exact matching between the $\theta$-term in the UV theory and the corresponding term in the effective theory, while the explicit matching of correlators in Ref.~\cite{Ai:2020ptm} shows that this assumption is invalid.

While we go with the assumption that the total derivative term $G^{\mu\nu}\widetilde{G}_{\mu\nu}$ can be traded for a scalar auxiliary field $q(x)$, presumably because of
nonperturbative quantum effects, $\int_\Omega \d^4 x\, q(x)$ (where $\Omega$ indicates the full volume) should still respect the global topological constraints. In particular, $\int_\Omega \d^4 x\, q(x)$ in each topological
sector should be exactly invariant under continuous deformations of the fields and therefore should also remain invariant when summing over sectors. Without further ado, using $q(x)$ as an unconstrained auxiliary field would hence clash with these global constraints because $q(x)$ would admit fluctuations that do not occur in the fundamental theory. In the effective chiral action one should therefore replace $\int_\Omega \d^4 x\, q(x)$ with some localized version that is not subject to topological constraints. This could be achieved for example by deriving an
effective theory in a large but finite subvolume $\Omega_1$ of the infinite volume $\Omega$. Such a theory can be obtained by integrating out fluctuations in the volume complement $\Omega_2\equiv \Omega\setminus\Omega_1$. In the subvolume $\Omega_1$ there are no physical constraints on the boundary conditions, and $\int_{\Omega_1}\d^4 x\, q(x)$ can fluctuate freely. Such an effective theory is derived in Ref.~\cite{Ai:2020ptm}, and it leads to correlators of operators with support in  $\Omega_1$ of the following form (in the general basis for the chiral phases):
\begin{align}
\label{exp:val:local}
\langle {\cal O}_1\rangle=&\frac{\sum\limits_{\Delta n_1=-\infty}^\infty \int\limits_{\Delta n_1} {\cal D}\phi\,{{(-1)^{-N_f\Delta n_1}{\rm e}^{-{\rm i} \, \bar\alpha \Delta n_1}}}{\cal O}_1\,{\rm e}^{-S_{\Omega_1}[\phi]}}{\sum\limits_{\Delta n_1=-\infty}^\infty \int\limits_{\Delta n_1} {\cal D}\phi\,{{(-1)^{-N_f\Delta n_1}{\rm e}^{-{\rm i}\,\bar\alpha    \Delta n_1}}}{\rm e}^{-S_{\Omega_1}[\phi]}}
\,,
\end{align}
where $\Delta n_1=\int_{\Omega_1} {\rm}d^4x\, q(x)$ is the topological charge within $\Omega_1$.
Here, $\theta$ is absent, and the phase factors $(-1)^{-N_f \Delta n_1}\e^{-\i \bar\alpha \Delta n_1}$ exactly cancel the $CP$ phases arising from the fermion determinants in $\Omega_1$ within each topological sector, resulting in no effects involving $\bar\theta$. One observes that the effective theory given in Eq.~\eqref{exp:val:local} instead contains a topological term $-{\rm i}\bar\alpha\Delta n_1$. Hence, while Eq.~\eqref{eq:Vecchia-Veneziano} with $\theta$ replaced by a general phase $\xi$ serves as an ansatz in which the phases are to be fixed by matching to the ’t~Hooft vertex, it follows that the choice $\xi=-\bar\alpha$ in Eq.~\eqref{eq:options2} is consistent with the topological term appearing in the Lagrangian of the subvolume effective theory. This effective theory accounts for quantum fluctuations in the complementary volume $\Omega_2$, providing an explicit example of how parameter values in an effective description may differ from those in the underlying fundamental theory, as discussed above.

{\section*{Acknowledgements}}
C.T. acknowledges conversations with J.~Benabou and support by the Cluster of Excellence “Precision Physics, Fundamental Interactions, and Structure of Matter” (PRISMA$^+$ EXC 2118/1) funded by the Deutsche Forschungsgemeinschaft (DFG, German Research Foundation) within the German Excellence Strategy (Project No. 390831469). \\

\bibliographystyle{utphys}
\bibliography{biblio}{}

@unpublished{2025quantum-rotor,
    author = {Ai, Wen-Yuan and Garbrecht, Bj\"orn and Tamarit, Carlos},
    title = "{The quantum rotator as an analogue model for topological effects in Yang--Mills theory}",
    note = "{\it in preparation}"
}

@article{DiVecchia:1980yfw,
    author = "Di Vecchia, P. and Veneziano, G.",
    title = "{Chiral Dynamics in the Large n Limit}",
    reportNumber = "CERN-TH-2814",
    doi = "10.1016/0550-3213(80)90370-3",
    journal = "Nucl. Phys. B",
    volume = "171",
    pages = "253--272",
    year = "1980"
}

@article{Benabou:2025viy,
    author = "Benabou, Joshua N. and Hook, Anson and Manzari, Claudio Andrea and Murayama, Hitoshi and Safdi, Benjamin R.",
    title = "{Clearing up the Strong $CP$ problem}",
    eprint = "2510.18951",
    archivePrefix = "arXiv",
    primaryClass = "hep-ph",
    month = "10",
    year = "2025"
}

@article{Ai:2024vfa,
    author = "Ai, Wen-Yuan and Garbrecht, Bjorn and Tamarit, Carlos",
    title = "{The QCD theta-parameter in canonical quantization}",
    eprint = "2403.00747",
    archivePrefix = "arXiv",
    primaryClass = "hep-th",
    reportNumber = "KCL-PH-TH/2024-13, TUM-HEP-1499/24, MITP-24-031",
    month = "3",
    year = "2024"
}

@article{Albandea:2024fui,
    author = "Albandea, David and Catumba, Guilherme and Ramos, Alberto",
    title = "{Strong CP problem in the quantum rotor}",
    eprint = "2402.17518",
    archivePrefix = "arXiv",
    primaryClass = "hep-lat",
    doi = "10.1103/PhysRevD.110.094512",
    journal = "Phys. Rev. D",
    volume = "110",
    number = "9",
    pages = "094512",
    year = "2024"
}

@article{Ai:2024cnp,
    author = "Ai, Wen-Yuan and Garbrecht, Bjorn and Tamarit, Carlos",
    title = "{$CP$ Conservation in the Strong Interactions}",
    eprint = "2404.16026",
    archivePrefix = "arXiv",
    primaryClass = "hep-ph",
    reportNumber = "KCL-PH-TH/2023-51, MITP-24-043, TUM-HEP-1508/24",
    doi = "10.3390/universe10050189",
    journal = "Universe",
    volume = "10",
    number = "5",
    pages = "189",
    year = "2024"
}

@article{Ai:2020ptm,
    author = {Ai, Wen-Yuan and Cruz, Juan S. and Garbrecht, Bj{\"o}rn and Tamarit, Carlos},
    title = "{Consequences of the order of the limit of infinite spacetime volume and the sum over topological sectors for $CP$ violation in the strong interactions}",
    eprint = "2001.07152",
    archivePrefix = "arXiv",
    primaryClass = "hep-th",
    reportNumber = "TUM-HEP-1249/20, CP3-20-02",
    doi = "10.1016/j.physletb.2021.136616",
    journal = "Phys. Lett. B",
    volume = "822",
    pages = "136616",
    year = "2021"
}

@article{Kaplan:2024ezz,
    author = "Kaplan, David B. and Sen, Srimoyee",
    title = "{Regulated chiral gauge theory and the strong CP problem}",
    eprint = "2412.02024",
    archivePrefix = "arXiv",
    primaryClass = "hep-lat",
    reportNumber = "INT-PUB-24-040",
    month = "12",
    year = "2024"
}

@article{Witten:1979vv,
    author = "Witten, Edward",
    title = "{Current Algebra Theorems for the U(1) Goldstone Boson}",
    reportNumber = "HUTP-79/A014",
    doi = "10.1016/0550-3213(79)90031-2",
    journal = "Nucl. Phys. B",
    volume = "156",
    pages = "269--283",
    year = "1979"
}

@article{Veneziano:1979ec,
    author = "Veneziano, G.",
    title = "{U(1) Without Instantons}",
    reportNumber = "CERN-TH-2651",
    doi = "10.1016/0550-3213(79)90332-8",
    journal = "Nucl. Phys. B",
    volume = "159",
    pages = "213--224",
    year = "1979"
}

@article{Nakamura:2021meh,
    author = "Nakamura, Y. and Schierholz, G.",
    title = "{The strong CP problem solved by itself due to long-distance vacuum effects}",
    eprint = "2106.11369",
    archivePrefix = "arXiv",
    primaryClass = "hep-ph",
    reportNumber = "DESY 21-078, DESY-21-078",
    doi = "10.1016/j.nuclphysb.2022.116063",
    journal = "Nucl. Phys. B",
    volume = "986",
    pages = "116063",
    year = "2023"
}

@article{Schierholz:2024var,
    author = "Schierholz, Gerrit",
    title = "{Absence of strong CP violation}",
    eprint = "2403.13508",
    archivePrefix = "arXiv",
    primaryClass = "hep-ph",
    reportNumber = "DESY-24-038",
    doi = "10.1088/1361-6471/adc31d",
    journal = "J. Phys. G",
    volume = "52",
    number = "4",
    pages = "04LT01",
    year = "2025"
}

@article{Schierholz:2025tns,
    author = "Schierholz, Gerrit",
    title = "{Absence of CP Violation in the Strong Interaction: Vacuum thwarts Axion}",
    eprint = "2502.04092",
    archivePrefix = "arXiv",
    primaryClass = "hep-lat",
    reportNumber = "DESY-25-21",
    doi = "10.22323/1.466.0398",
    journal = "PoS",
    volume = "LATTICE2024",
    pages = "398",
    year = "2025"
}

@book{Srednicki:2007qs,
    author = "Srednicki, M.",
    title = "{Quantum field theory}",
    doi = "10.1017/CBO9780511813917",
    isbn = "978-0-521-86449-7, 978-0-511-26720-8",
    publisher = "Cambridge University Press",
    month = "1",
    year = "2007"
}

@article{Shifman:1979if,
    author = "Shifman, Mikhail A. and Vainshtein, A. I. and Zakharov, Valentin I.",
    title = "{Can Confinement Ensure Natural CP Invariance of Strong Interactions?}",
    reportNumber = "ITEP-64-1979",
    doi = "10.1016/0550-3213(80)90209-6",
    journal = "Nucl. Phys. B",
    volume = "166",
    pages = "493--506",
    year = "1980"
}
  
\end{document}